# Metal-to-Insulator Switching in Quantum Anomalous Hall States

Xufeng Kou,[1,†] Lei Pan,[1,†] Jing Wang,[2] Yabin Fan,[1] Eun Sang Choi,[3] Wei-Li Lee,[4] Tianxiao Nie,[1] Koichi Murata,[1] Qiming Shao,[1] Shou-Cheng Zhang,[2] and Kang L. Wang [1*]

[1]Device Research Laboratory, Department of Electrical Engineering, University of California, Los Angeles, California, USA, 90095

[2]Department of Physics, Stanford University, Stanford, California, USA, 94305

[3]National High Magnetic Field Laboratory, Tallahassee, Florida, USA, 32310

[4]Institute of Physics, Academia Sinica, Taipei, Taiwan, 11529

†These authors contributed equally to this work.

*To whom correspondence should be addressed. E-mail: wang@seas.ucla.edu

**After decades of searching for the dissipationless transport in the absence of any external magnetic field, quantum anomalous Hall effect (QAHE) was recently achieved in magnetic topological insulator (TI) films. However, the universal phase diagram of QAHE and its relation with quantum Hall effect (QHE) remain to be investigated. Here, we report the experimental observation of the giant longitudinal resistance peak and zero Hall conductance plateau at the coercive field in the 6 quintuple-layer $(Cr_{0.12}Bi_{0.26}Sb_{0.62})_2Te_3$ film, and demonstrate the metal-to-insulator switching between two opposite QAHE**

**plateau states up to 0.3 K. Moreover, the universal QAHE phase diagram is realized through the angle-dependent measurements. Our results address that the quantum phase transitions in both QAHE and QHE regimes are in the same universality class, yet the microscopic details are different. In addition, the realization of the QAHE insulating state unveils new ways to explore quantum phase-related physics and applications.**

When a two-dimensional electron gas (2DEG) is subjected to a strong perpendicular magnetic field ($B_\perp$), the energy spectrum evolves into discrete Landau levels (LLs). As a result, the electron motion is localized by the cyclotron orbits inside the 2DEG system, while the one dimensional chiral states are formed at the edge, thus giving rise to a quantized Hall conductance[1-3]. Ever since the discovery of the quantum Hall effect (QHE)[4], enormous efforts have been made to elucidate the universal behavior of this quantum transport phenomenon[5-7]. Within the framework of 2D localization theorem, the renormalization group (RG) flow of the system can be well-described in the conductance plot, where stable points appear at ($\sigma_{xx}$, $\sigma_{xy}$) = (0, $ne^2/h$) ($\sigma_{xx}$ is the longitudinal conductance, $\sigma_{xy}$ is the transverse conductance, $e$ is the electron charge, $h$ is the Planck constant, and the integer $n$ is in accordance with the $n^{th}$ LL.). In the meanwhile, the quantum phase transition between neighboring LLs follows a continuous semicircle centered at ($\sigma_{xx}$, $\sigma_{xy}$) = (0, $(n-1/2)e^2/h$) with the scaling behavior dictated by the localization length[8-11]. In the extreme quantum limit when $B_\perp$ is so large that the ground LL exceeds the Fermi level ($E_F$), the 2DEG system will be driven into the so-called quantum Hall insulator state where the longitudinal resistance diverges as the temperature approaches absolute zero, yet the Hall resistance saturates towards the quantized value corresponding to the ground LL state[7, 12-14].

Following the footprint of QHE, it was suggested that by introducing appropriate magnetic exchange field and large intrinsic spin-orbit coupling, similar dissipationless chiral edge conduction could also occur in non-zero first Chern number ($C_1$) ferromagnetic (FM) insulators without the assistance of external magnetic field[15-21]. Given the same broken symmetry and non-trivial band topology, the corresponding quantum anomalous Hall effect (QAHE) may be regarded as a special derivative of QHE[17, 22, 23]. Accordingly, soon after the first experimental demonstrations of QAHE in magnetic topological insulators (TIs)[24-28], a microscopic network model of quantum percolation was applied to investigate the universal scaling of the QAHE plateau transition around the coercive field[29]. Generally, it was proposed that in an ideal 2D magnetic TI system, the competition between the FM exchange field and the quantum confinement-induced hybridization would change the band topology, therefore leading to the phase transition from the $C_1 = 1$ QAHE state to the $C_1 = 0$ insulating state. Unfortunately, due to the major challenge of growing high quality magnetic TIs (i.e., well-defined single-crystalline structure with atomically smooth surfaces and extremely low defect density) within the 2D hybridization limit[24-26], experimental observations of the zero Hall plateau state in a quantum anomalous Hall insulator and the plateau transition from $\sigma_{xy} = \pm e^2/h$ to $\sigma_{xy} = 0$ in the QAHE regime have not been reported yet. Meanwhile, although there was an earlier attempt to interpret QAHE following the RG flow argument[26], the universality of this distinct QAHE phase remains to be ascertained.

**Results**

**QAHE in the Cr-doped $(BiSb)_2Te_3$ sample**

In this article, we use the Cr-doped $(BiSb)_2Te_3$ films grown on semi-insulating GaAs (111)B

substrates via molecular beam epitaxy (MBE) to study QAHE and its related phase diagram[30]. The growth condition and chemical composition of the film were carefully optimized such that extremely low bulk carrier density was achieved and the Fermi level was already within the surface gap without additional electric field tuning (See Supplementary Note 1). In order to generate the hybridization surface state gap $m_0$ [19, 31, 32], the film thickness used in this work was precisely chosen as 6 quintuple-layer (QL), as highlighted in Supplementary Fig. 1. After careful in-situ surface passivation[33], we carried out standard four-point magneto-transport measurements on the 6 QL $(Cr_{0.12}Bi_{0.26}Sb_{0.62})_2Te_3$ Hall bar device with dimensions of 2 mm × 1 mm, as illustrated in Fig. 1a. Figure 1b shows the quantization of the anomalous Hall resistance ($R_{yx} = h/e^2 \sim 25.81$ k$\Omega$) up to $T = 0.26$ K. The sign of the quantized $R_{yx}$ with respect to the magnetization direction is consistent with the chiral conduction property, as elaborated in ref. 25. Likewise, the magnetic field-dependent longitudinal resistance ($R_{xx}$) results are plotted in Fig. 1c (The complete temperature-dependent data of $R_{yx}$ and $R_{xx}$ are provided in Supplementary Fig. 3). The striking feature here is that at the base temperature $T = 0.02$ K, $R_{xx}$ increases dramatically from almost a vanishing value ($R_{xx\_min} \sim 20$ $\Omega$) at zero field to a giant peak ($R_{xx\_PEAK} \sim 380$ k$\Omega$) at the coercive field ($H_C = \pm 0.12$ T); both $R_{xx\_PEAK}$ and the corresponding magneto-resistance (MR) ratio (> $10^6$ %) are much larger than those reported previously[24-28]. It is noted that the steep divergence of $R_{xx}$ at zero LL was also observed in graphene under high magnetic field[34-36]. However, the underlying mechanisms are quite different. For the graphene case, it was believed that the increased exchange interaction at high $B_\perp$ would split the spin degeneracy of graphene around the Dirac point, and lift the original fourfold degenerate Dirac cone by a Zeeman gap[34-38]. In contrast, both the giant $R_{xx\_PEAK}$ at small $H_C$ and its temperature-dependent behavior shown in our 2D magnetic TI film (Fig. 1d and Supplementary Fig. 2) can be understood from the magnetic multi-domain network model at the QAHE phase transition regime, as we will explain in detail below.

**Capturing the $e^2/h$-to-zero QAHE plateau transition**

Following ref. 29, we re-plot the QAHE data on the basis of conductance in Fig. 2 by using the reciprocal resistance-to-conductance ($\rho$ - $\sigma$) tensor conversion given by

$$\sigma_{xx} = \frac{\rho_{xx}}{\rho_{xx}^2 + \rho_{yx}^2}, \quad \sigma_{xy} = \frac{\rho_{yx}}{\rho_{xx}^2 + \rho_{yx}^2} \quad (1)$$

Consistent with Figs. 1b-c, the QAHE in the conductance plot is also manifested by the quantized ($\sigma_{xy} = \pm e^2/h$, $\sigma_{xx} \to 0$) at zero magnetic field, as shown in Figs. 2a-b. Remarkably, two intermediate plateaus with the Hall conductance $\sigma_{xy} \sim 0$ are clearly developed at $H_C = \pm 0.12$ T when $T = 0.02$ K. In the meantime, the longitudinal conductance $\sigma_{xx}$ exhibits the double-peaks behavior, and its minimum value at $H_C$ reflects the insulating behavior during the magnetization reversal process. Moreover, the temperature dependence of the zero Hall plateaus is also investigated. Figs. 2a - b present the measured $\sigma_{xy}$ and $\sigma_{xx}$ data at several different temperatures (i.e., 0.02 K to 0.33 K). It is seen that with increasing temperature, the zero Hall plateaus near $H_C$ gradually narrow (Fig. 2a), while the $\sigma_{xx}$ double-peaks widen at high temperatures, and the minimum of $\sigma_{xx}$ at $H_C$ also gradually becomes larger, indicating the increase of thermally activated bulk carrier conduction at higher temperatures (Fig. 2b). When the sample further warms up, both these intermediate features cannot be resolved any more at $T = 1.9$ K.

To further validate our observations, we investigate more Cr-doped (BiSb)$_2$Te$_3$ thin films with different thickness (6 and 10 QL, see Supplementary Note 5) and Cr doping concentrations (10% to 15%, see Supplementary Note 6), and the relevant transport data are shown in Supplementary Figs. 5 and 6, respectively. It is seen that while the quantized QAHE states are observed in all samples, the zero Hall plateau state only persists in the 6 QL films and disappears when the thickness increases to 10 QL. Together with Fig. 2, these thickness-dependent features strongly suggest that the signatures of physics behind the quantized QAHE state ($\sigma_{xy} = \pm e^2/h$, $\sigma_{xx} \rightarrow 0$) and the zero Hall plateaus ($\sigma_{xy} \sim 0$, $\sigma_{xx} \rightarrow 0$) are different, and they may be understood as the following based on the mean field theory[29]. In principle, for a 2D magnetic TI system, there are two mechanisms to open the surface state gap: one is the hybridization gap $m_0$ between the top and bottom surface states due to quantum confinement, and the other is the exchange field gap $\Delta_M$ introduced by the FM ordering along $z$-direction[19, 29, 39]. In the quantized QAHE state with $|\Delta_M| > |m_0|$, all the magnetic domains in the magnetic TI film are well-aligned along the same direction, and the corresponding first Chern number of the system is $C_1 = \Delta_M/|\Delta_M| = \pm 1$. Consequently, there is only one single chiral edge state propagating along the sample edge so that ($\sigma_{xy} = \pm e^2/h$, $\sigma_{xx} \rightarrow 0$). On the other hand, the zero Hall plateau occurs around the coercive field during the magnetization reversal process, where many upward and downward domains coexist in a random manner[40, 41], as schematically shown in Fig. 2c. Under the mean field approximation, now $|\Delta_M| < |m_0|$ and the first Chern number of the system is $C_1 = 0$. Accordingly, $\sigma_{xy}$ develops the zero plateaus in the $-|m_0| < \Delta_M < |m_0|$ regime, and $\sigma_{xx}$ approaches zero.

Recently, some of the authors proposed a microscopic network model to describe the critical behavior of the QAHE plateau transition in the magnetic TI system[29]. It was suggested that zero plateaus in $\sigma_{xy}$ could occur and $\sigma_{xx}$ would show two peaks around $H_C$. The basic physical picture is that, at the coercive field, the chiral edge states that are located at the magnetic domain boundaries may tunnel into each other when the spatial decay length of edge states is larger than the distance between them. Under such circumstances, the QAHE plateau transition at $H_C$ can be mapped into the network model of the integer QHE plateau transition in the lowest LL. Therefore, the phase transition from the quantized QAHE state to the zero Hall plateau should show critical properties, where the universal temperature and size scaling behavior of $\sigma_{xy}$ and $\sigma_{xx}$ were proposed[29].

Following the above suggestion[29], temperature-dependent $\sigma_{xy}$ slope $S = (\partial\sigma_{xy}/\partial H)_{max}$ is displayed in Fig. 2d to study the scaling behavior of the QAHE plateau transition. Although $S$ is found to monotonically decrease versus temperature (which is consistent with the theoretical proposal), we should point out that there are some quantitative differences between the theoretical proposal and our experimental results. First, unlike the simulation results[29], both zero $\sigma_{xy}$ plateaus and double-split $\sigma_{xx}$ persist even when the system has already deviated from the perfect quantization case at 0.33 K (i.e., $\sigma_{xy}$ = 0.992 $e^2/h$ and $\sigma_{xx}$ picks up a relative large background signal of 0.127 $e^2/h$). Second, the temperature–dependence of $S$ seems not to follow the predicted simple $S \propto T^{-\kappa}$ relation in the entire temperature range (0.02 K to 0.33 K); yet when $T < 0.1$K, the power law scaling is fitted quite well with $\kappa = 0.22$, which is just one half of $\kappa = 0.42$ measured in the QHE transition[10]. The possible reasons for the above differences may be due to the fact that the coercivity shifts when 0.1K $< T <$ 0.33 K compared to $T <$ 0.1 K (See Supplementary Note 3), and extra thermally activated bulk carriers would also contribute to the transport (i.e., $R_{xx}$ increases dramatically when $T >$ 0.1K, as highlighted in Supplementary Note 2 and the Inset of Supplementary Fig. 2b)[27], therefore complicating the slope of $\sigma_{xy}$ in high temperature regime (Also, the complicated magnetic domain dynamics at the coercive field may also affect the temperature scaling[41], which is not included in the original network model for QHE plateau transition). More importantly, it is noted that $\kappa$ is not a universal exponent; instead, it strongly depends on the microscopic details of the randomness in magnetic domains[29]. Specifically, $\kappa$ is defined to be $p/2\nu$, where $p$ is the exponent of temperature dependence of the inelastic scattering length with $L_{in} \propto T^{-p/2}$, and $v$ is the real universal critical exponent[29]. For QHE systems, $v$ is found to be a constant of 2.4, $p = 2$ for high-mobility sample and $p = 1$ for "dirty sample"[42]. In our case, by assuming the same $v = 2.4$, we obtain $p = 1$ which seems to be consistent with our sample property (i.e., low carrier mobility[25]). Nevertheless, the exact values of both $v$ and $p$ in the QAHE systems remain to be further investigated.

Finally, we would like to point out that the insulating state in the 2D QAHE case is different from the quantum Hall insulator in conventional 2DEG systems in terms of resistivity. In particular, as $T \to 0$, the zero-Hall plateau QAHE insulator approaches ($\rho_{xx} \to \infty$, $\rho_{yx} \to 0$), while the QHE insulator exhibits ($\rho_{xx} \to \infty$, $\rho_{yx} \to vh/e^2$) where $v$ is the lowest LL filing factor[7,14]. This is due to the differences of the band structures: for magnetic TI, the linear Dirac-cone-like surface states enable the Fermi level to be located at the Dirac point; yet for ordinary 2DEG with parabolic energy dispersion relation, no zero LL is allowed, and the minimum value of $\rho_{yx}$ thus has to saturate at a non-zero ground quantized value of $vh/e^2$. Even for QHE in 2D Dirac fermion system (i.e., graphene) with zero[th] LL[43], the observed zero-energy state ($\rho_{xx} \to \infty$, $\rho_{yx} \to 0$) at high magnetic field[34-36] is microscopically different from the zero-Hall plateau QAHE insulator discussed above. Although the measured "zero-Hall plateaus" in Supplementary Fig. 4 exhibit a flat linear slope with the values of $\sigma_{xy}$ below 1 μS (since experimentally $R_{xx\_PEAK}$ is always finite), the $e^2/h$-to-zero QAHE plateau transition is still manifested by the dramatic change of the slope of $\sigma_{xy}$ (see Supplementary Note 4). Moreover, extra transport data in other 6 QL samples all show the zero Hall plateaus around $H_C$ (See Supplementary Note 6), again suggesting the quantum phase transition between the $C_1 = 1$ QAHE state and the $C_1 = 0$ insulating state in the 2D hybridization regime.

**Mapping the semicircle QAHE phase diagram**

In light of the importance of QAHE phase transition, we further performed the angle-dependent measurements. It has been shown the Cr-doped magnetic TI develops robust out-of-plane FM order in the QAHE regime[19, 30, 44]. If the applied magnetic field is not perfectly perpendicular to the film, the induced in-plane magnetic component tilts the Cr magnetization, and thereby helps to modulate the strength of $\varDelta_M$ as well. Figures 3a - b show the magneto-transport results of the 6 QL $(Cr_{0.12}Bi_{0.26}Sb_{0.62})_2Te_3$ film as the tilted angle $\theta$ of the Hall-bar device with respect to the magnetic field is varied from 90° (out-of-plane) to 180° (in-plane). We observe that with small magnetic field sweeping (-0.5 T < $B$ < 0.5 T) at $T$ = 0.02 K, the quantization of $R_{yx}$ (Fig. 3a) and the zero Hall plateau (Supplementary Fig. 7) are relatively robust as long as $\theta$ < 150°, as addressed in Supplementary Note 7. On the other hand, as both $B$ and $\theta$ increase, the system starts to deviate from the QAHE state. In the extreme case when the film is rotated almost parallel with the magnetic field ($\theta$ = 180° ± 5°), the measured in-plane magneto-resistance reflects a rather insulating feature: the corresponding dashed purple curve in Fig. 3b strongly suggests the system is rapidly approaching toward an angle-induced insulating state with $R_{xx}$ > 400 kΩ. Note that due to the giant perpendicular anisotropy in the Cr-doped TI systems, out-of-plane magnetic domains might still be formed in this case when $\theta$ ~ 180°[45-47]. A more detailed experiment was carried out subsequently in which $R_{xx}$ and $R_{yx}$ were recorded when the 6 QL magnetic TI film was continuously rotated from 90° to 180°, under different fixed applied magnetic fields. From Figs. 3c - d, it is clear that the system undergoes the smooth quantum phase transition with respect to $\theta$. Most importantly, we find that both $R_{xx}$ and $R_{yx}$ curves for $B$ ≥ 1 T tend to change between points [$R_{xx}$(90°) = 0, $R_{xx}$(180°) ~ 15 $h/e^2$] and [$R_{xy}$(90°) = $h/e^2$, $R_{yx}$(180°) = 0], while different magnetic fields only modulate the transition process in between: $R_{yx}$ rolls off the $h/e^2$ quantization line more quickly and the divergence of $R_{xx}$ occurs at smaller $\theta$ when **B** increases.

We can further visualize the angle-assisted QAHE phase transition in Fig. 4. Significantly, when displayed in the $\sigma_{xy}$ - $\sigma_{xx}$ plot, all $\theta$-dependent curves for $B \geq 1T$, which have been manually shifted vertically by $(B-1)\times 0.1 e^2/h$ for comparison in Fig. 4a, follow a single continuous semicircle which is centered at $(\sigma_{xx}, \sigma_{xy}) = (0, e^2/2h)$ with the radius of $e^2/2h$. Meanwhile, another interesting finding in Figs. 4a - c is the importance of the in-plane magnetic field $B_{//} = B \cdot \cos\theta$ on the universal QAHE phase diagram. If we divide the conductance semicircle into three angle regions (red dots for [90°, 120°], green dots for [90°, 150°], and blue dots for [150°, 180°]), and track the evolutions of each component versus the applied magnetic field, we see that it is the strength of $B_{//}$ rather than the $B_{//} / B_{\perp}$ ratio that determines the QAHE phase transition. As highlighted by the dashed curves in Fig. 4a, it is evident that even though $B_{//} / B_{\perp} = \cot\theta$ is always smaller than 1 within [90°, 120°], the in-plane $B_{//}$ under high magnetic fields ($B > $ 1T) still manage to force the conduction deviate from the dissipationless $(0, h/e^2)$ point, and such spread-out trend becomes more pronounced with increased **B** (and thus $B_{//}$). In contrast, as long as the total magnetic field **B** is smaller than the critical magnetic field ($B_0 \sim 1$ T), the semicircle phase transition cannot be completed even when $\theta = 180°$ ($B_{//,max} = B$), as manifested in Fig. 4b. Finally, by combining both the field-dependent results of Fig. 2 and the angle-dependent data of Fig. 4a ($\theta \subseteq [90°, 270°]$, $B = 1$ T) together in the $\sigma_{xy}$ - $\sigma_{xx}$ plot, we produce a single semicircle curve in Fig. 4d, illustrating the similar QAHE phase transition feature between the $C_1 = 0$ zero Hall plateau state and the two $C_1 = \pm 1$ QAHE states. Given that the scaling rule of localization in the multi-domain configuration is temperature-dependent[29], the deviation from the two QAHE-characterized $(0, 0)$ and $(0, e^2/h)$ points is indeed found to become more obvious with increasing $T$, as shown in Fig. 4e and Supplementary Fig. 8(Relevant explanations are given in Supplementary Note 8).

**Discussions**

Conductance semicircles similar to those discussed in above section have been extensively investigated to describe the global phase diagram of QHE[5-8, 13, 14], yet we emphasize here that the microscopic physics between the QHE and QAHE phase transitions are different. Specifically speaking, the QHE phase diagram is closely related to the LL quantization. The corresponding quantum Hall plateau transitions happen when the Fermi level crosses mobility edges, which are due to disorder-induced localization-delocalization transitions[6]. Concurrently, the quantum Hall insulator is achieved when the applied $B_\perp$ is large enough to drive the ground LL overlaps with $E_F$ [14, 35]. On the contrary, in the QAHE state, the zero quantized Hall plateau may be the result of the multi-domain formation and a network of chiral edge states at domain walls during the magnetization reversal process[29]. As a result, the phase transition to the zero Hall plateau state in the QAHE regime can be obtained at a much smaller magnetic field, as addressed in Figs. 2 and 4. Furthermore, since the first Chern number in the QAHE state is determined by the competition between magnetic exchange gap and hybridization gap[29], it is thus suggested that in 3D magnetic TIs where higher subbands may participate into the band topology transition[48], new QAHE phases with tunable $C_1$ are expected, and relevant phase transition can thus be further modulated by film thickness.

In conclusion, we study the QAHE phase transition for the 2D hybridized magnetic TI system. We show that such QAH metal-to-insulator switching can only be achieved in high-quality samples with truly bulk insulating state and 2D quantum confinement. The observations of the zero Hall plateaus and double-split longitudinal conductance are consistent with the proposed microscopic multi-domain network model where the vanishing of the magnetic exchange gap $\Delta_M$ at the coercive field causes the topology change, yet the temperature-dependent and size scaling behaviors of the QAHE plateau transition needs further investigations to reveal the nature of this exotic state of matter. At the same time, from both the field-dependent and angle-dependent magneto-transport results, we map out the global QAHE phase diagram which can be described by a single semicircle curve continuously connecting the (0, 0) and (0, $e^2/h$) in the ($\sigma_{xx}$, $\sigma_{xy}$) conductance plot. Additionally, we achieve the QAHE insulator regime by making either ($B = \pm \mu_0 \cdot H_C$, $\theta = 90°$) or ($B > B_0$, $\theta = 180°$) at relatively small magnetic fields. The discovered universal phase transition rule is significant for the understanding of the QAHE system and our results open new avenues for the exploration of novel QAHE-related phenomena and applications.

**Note added**. During the preparation of the manuscript, we are aware of a related work by Y. Feng et al.[49] that reports the observation of the zero Hall plateau (although the sample is not fully quantized) in a back-gated 5 QL Cr-doped magnetic TI sample.

**Methods**

**MBE growth.** High-quality single crystalline Cr-doped ($Bi_xSb_{1-x}$)$_2$Te$_3$ films were performed in an ultra-high vacuum Perkin-Elmer MBE system. Semi-insulating ($\rho > 10^6$ $\Omega \cdot$cm) GaAs (111)B substrates were cleaned by acetone with ultrasonic for 10 minutes before loaded into the growth chamber. Then the substrates were annealed to 580 °C to remove the native oxide, under Se rich environment. During the growth, the GaAs substrate was maintained around 200 °C (growth temperature), with the Bi, Sb, Te, and Cr shutters opened at the same time. Epitaxial growth was monitored by an in-situ RHEED technique, where the digital RHEED images were captured using a KSA400 system built by K-space Associates, Inc. After the film growth, a 2 nm Al was evaporated to passivate the surface at room temperature.

**Characterizations.** For magneto-transport measurements at 0.02 K < $T$ < 0.33 K, we used a $He_3/He_4$ dilution refrigerator system equipped with a 18-Tesla superconducting magnet at the SCM1 cell in NHMFL, Tallahassee, USA. For transport measurements at 1.9 K < $T$ < 300 K, we used the Quantum Design physical property measurement system (ppms). We are able to systematically alter several experimental variables such as temperature, magnetic field, working frequency and rotation angle. Multiple lock-in-amplifiers, Lake Shore AC resistance bridges, and Keithley source meters are also connected with the physical property measurement system, enabling comprehensive and high-sensitivity transport measurements for the Hall bar devices.

**Acknowledgements:**

We thank Prof. D. Goldhabor-Gordon for helpful discussion and collaboration. We are grateful to the support from the DARPA Meso program under contract No.N66001-12-1-4034 and N66001-11-1-4105, and the ARO program under contract No.W991NF-14-1-0607. We also acknowledge the support from the Western Institute of Nanoelectronics (WIN) and the support from the FAME Center, one of six centers of STARnet, a Semiconductor Research Corporation program sponsored by MARCO and DARPA. A portion of this work was performed at the National High Magnetic Field Laboratory, which is supported by National Science Foundation Cooperative Agreement No.DMR-1157490 and the State of Florida. K.L.W acknowledges the support of the Raytheon endorsement. J. W. and S. C. Z. also acknowledge the support from the U.S. Department of Energy, Office of Basic Energy Sciences, Division of Materials Sciences and Engineering, under contract No. DE-AC02-76SF00515. W.L.L. acknowledges funding support from the Academia Sinica 2012 career development award in Taiwan.

**Author Contributions**

X. K. and K. L. W. conceived and designed the research. X. K. and L. P. grew the material, X. K., L. P., and E. C. performed the measurements. Y. F., T. N., K. M., and Q. S. contributed to the measurements and analysis. J. W. L. L and S. Z. designed the theoretical model. X. K., J.W., and K.L. W. wrote the paper with help from all of the other co-authors.

**Additional information**

**Supplementary Information** accompanies this paper at http://www.nature.com/naturecommunications

**Competing financial interests**: The authors declare no competing financial interests.

**Reprints and permission** information is available online at http://npg.nature.com/reprintsandpermissions/

**How to cite this article**: Kou, X. *et al.* Metal-to-Insulator Switching in Quantum Anomalous Hall States. *Nat. Commun.*

**Figure captions:**

**Figure 1 | Quantum anomalous Hall effect in the 6 QL $(Cr_{0.12}Bi_{0.26}Sb_{0.62})_2Te_3$ film. a.** Schematic of the mm-sized Hall bar structure and four-point Hall measurements based on the MBE-grown magnetic TI thin film. **b.** Quantum anomalous Hall results at $T = 0.26$ K. The Hall resistances are quantized to be $\pm h/e^2$ where the signs are determined by the chirality of the edge conduction. **c.** Temperature-dependent magneto-resistance results (from 0.02 K to 1.9 K). At the coercive field, the peak of $R_{xx}$ quickly diverges at lower temperatures. **d.** Temperature-dependent $R_{xx\_PEAK}$ and $R_{yx}$ extracted from Figs. 1b - c. The anomalous Hall resistance $R_{yx}$ becomes quantized up to 0.3 K, and the giant $R_{xx\_PEAK}$ resolved at 0.02 K is around 400 kΩ, the largest value obtained among all reported QAHE systems.

**Figure 2 | Quantum phase transition of quantum anomalous Hall effect. a.** Magnetic field dependent $\sigma_{xy}$ at different temperatures. Zero Hall plateaus at $\pm H_C$ are developed between the two QAHE states up to 0.33 K. **b.** Magnetic field dependent $\sigma_{xx}$ at different temperatures. Even when the film already deviates from the perfect QAHE state at $T = 0.33$ K, both the zero $\sigma_{xy}$ plateaus and double-peaked $\sigma_{xx}$ still persist. **c.** Schematic of the multi-domain network formed during the magnetization reversal process. The downward green arrows and upward yellow arrows denote the up ($\Delta_M > |m_0|$) and down ($\Delta_M < -|m_0|$) magnetic domains. **d.** Temperature-dependent transition slope $S = (\partial \sigma_{xy} / \partial H)_{max}$ extracted from Fig. 2a. When $T < 0.1$ K, $S \propto T^{-\kappa}$ follows the power low scaling behavior with $\kappa = 0.22$.

**Figure 3 | Angle-dependent transport measurements in the quantum anomalous Hall regime. a.** Quantum anomalous Hall results when the 6 QL magnetic TI film is rotated to $\theta = $ 90°, 120°, and 150°, respectively. The external magnetic field is swept between -1 T and +1 T.

The sample temperature is 0.02 K. **b.** Magneto-resistance results of the sample under different tilted angle $\theta$ at $T = 0.02$ K. The film enters the QAHE insulating state when it is rotated parallel to the magnetic field. Angle-dependent **c.** $R_{yx}$ and **d.** $R_{xx}$ under different fixed $B$ at $T = 0.02$ K. All curves tend to converge at two critical points at $(R_{xx}, R_{yx}) = (0, h/e^2)$ and $(\sim 15\ h/e^2, 0)$.

**Figure 4 | Global phase diagram of quantum anomalous Hall effect. a.** Angle-assisted QAHE phase transition in the ($\sigma_{xx}$, $\sigma_{xy}$) plot. The 6 QL Cr-doped TI film is continuously rotated from 90°(out-of-plane) to 180° (in-plane) with respect to the applied magnetic field direction. All curves overlap with each other following the single semicircle relation. Data are shifted vertically by $(B-1) \times 0.1 e^2/h$ for convenient comparison. **b.** and **c.** Magnetic field dependent QAHE phase diagram in the 3D plot. When $B_{//} < 0.5$ T, the semicircle transition cannot be completed. **d.** Comparison between the field-dependent $R_{xx}$ - $R_{xy}$ results (with the sweeping direction of the applied magnetic field from +0.3 T to -0.3 T) and the angle-dependent data in Fig. 4a ($\theta \subseteq$ [90°, 270°] and $B = 1$ T). Identical QAHE phase transition between the $C_1 = 0$ insulating state and the two $C_1 = \pm 1$ QAHE states is manifested. **e.** Temperature-dependent ($\sigma_{xx}$, $\sigma_{xy}$) plot. As $T$ increases, the intrinsic localization scaling rule diverts the system from the original QAHE and insulating states. The dotted lines link the data points with the same angles ($\theta \subseteq$ [90°, 270°] with each step of 10°) between 0.02 K and 1.9 K.

**FIGURE LEGEND**

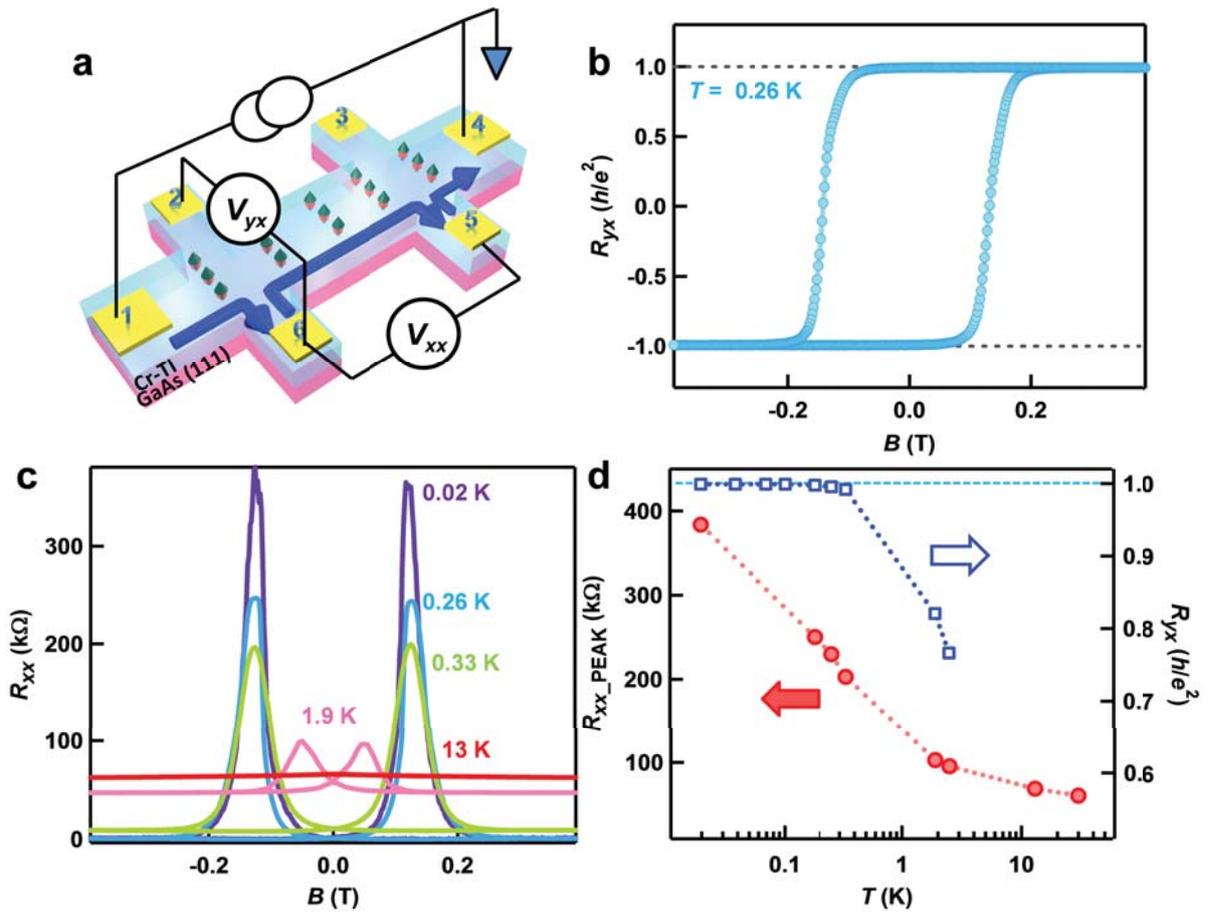

**Figure 1**

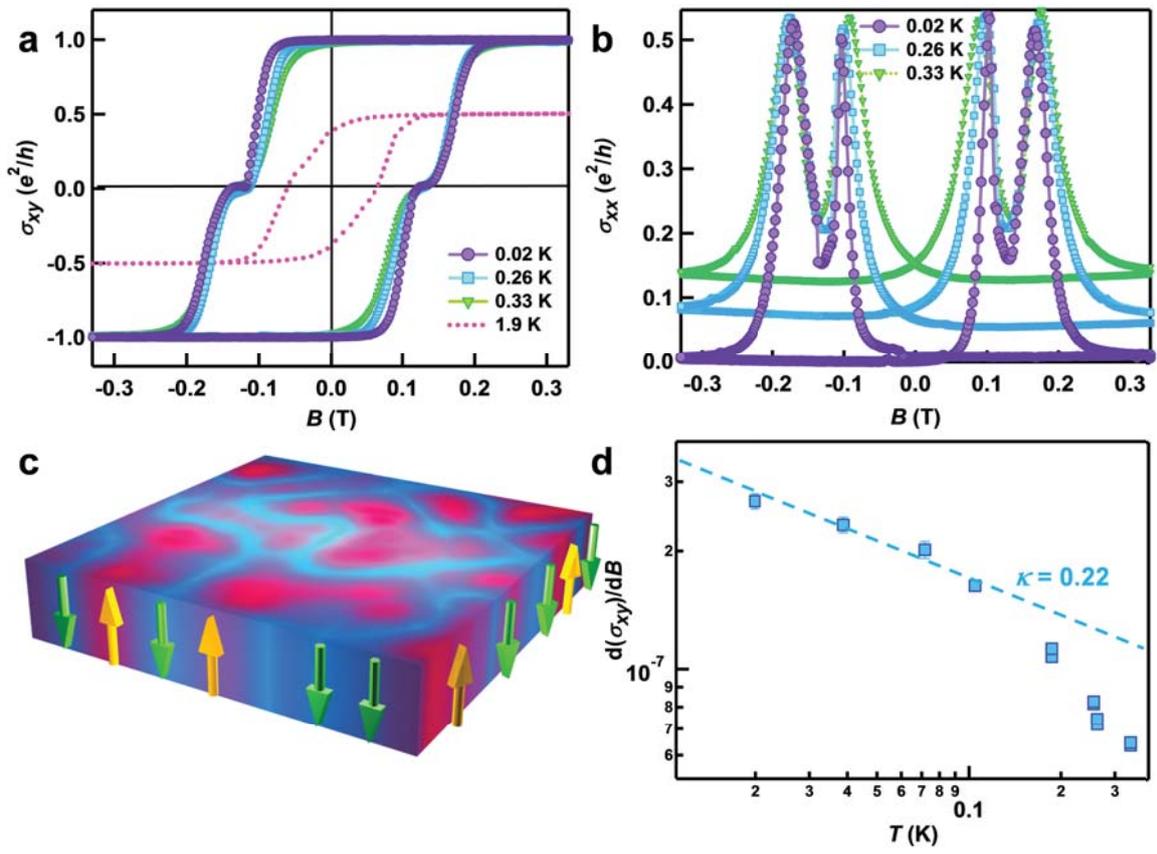

**Figure 2**

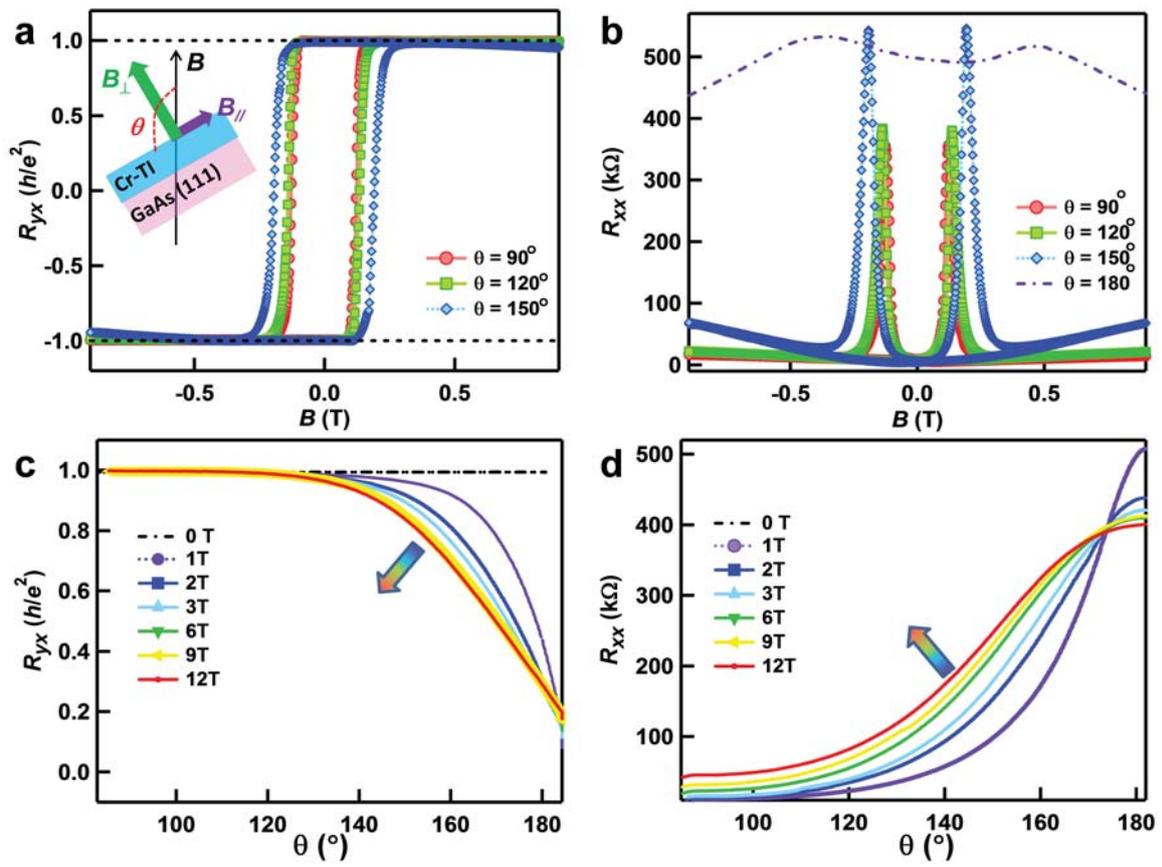

**Figure 3**

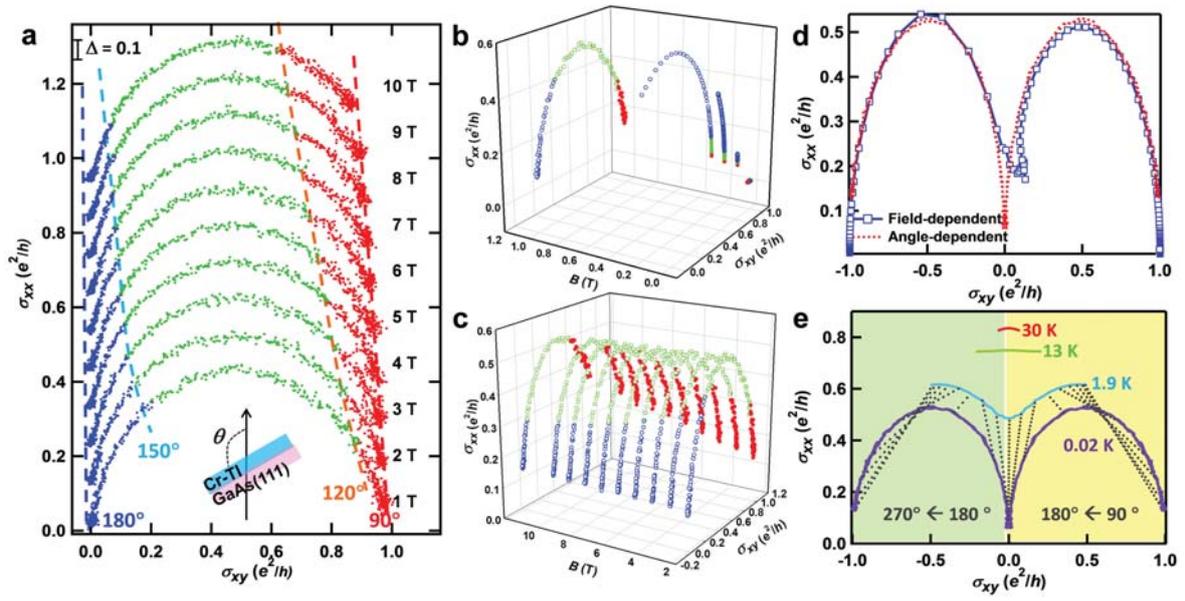

**Figure 4**